\documentclass[prl,superscriptaddress,groupaddress,letterpaper,twocolumn,pacs,nofootinbib]{revtex4-1}

\usepackage{natbib}
\usepackage{cancel}
\usepackage{amsthm}         
\usepackage{amsmath} 	   
\usepackage{amssymb}       
\usepackage{amsfonts}
\usepackage{graphicx} 	    
\usepackage{verbatim}
\usepackage{color,colortbl}
\usepackage{float}

\definecolor{dark-red}{rgb}{0.0,0.0,0.0}
\definecolor{dark-green}{rgb}{0.0,0.0,0.0}
\definecolor{dark-blue}{rgb}{0.0,0.0,0.0}

\definecolor{LightCyan}{rgb}{1,1,0.75}
\definecolor{LightGreen}{rgb}{0.8,1,.85}
\definecolor{LightRed}{rgb}{1,.75,.75}
\definecolor{PaleSilver}{rgb}{.83,.83,.83}
\definecolor{crazy}{rgb}{.9,.9,1}

\usepackage{enumitem}
\usepackage{multirow}
\usepackage[table]{xcolor}

\begin{document}

\title{Inflationary schism after Planck2013} 

\author{Anna Ijjas}
\affiliation{Max-Planck-Institute for Gravitational Physics (Albert-Einstein-Institute), 14476 Potsdam, Germany} \affiliation{Rutgers University, New Brunswick, NJ 08901, USA}

\author{Paul J. Steinhardt}
\affiliation{Department of Physics and Princeton Center for Theoretical Science, Princeton University, Princeton, NJ 08544, USA}

\author{Abraham Loeb}
\affiliation{Harvard-Smithsonian Center for Astrophysics, Cambridge, MA 02138, USA}

\date{\today}

\begin{abstract}
Classic inflation, the theory described in textbooks, is based on the idea that, beginning from typical initial conditions and assuming a simple inflaton potential with a minimum of fine-tuning, inflation can create exponentially large volumes of space that are generically homogeneous, isotropic and flat, with nearly scale-invariant spectra of density and gravitational wave fluctuations that are adiabatic, Gaussian and have generic predictable properties.   In a recent paper, we showed that, in addition to having certain conceptual problems known for decades, classic inflation is for the first time also disfavored by data, specifically the most recent data from WMAP, ACT and Planck2013.  Guth, Kaiser and Nomura and Linde have each recently published critiques of our paper, but, as made clear here, we all agree about one thing:  the problematic state of classic inflation.  Instead, they describe an alternative inflationary paradigm that revises the assumptions and goals of inflation, and perhaps of science generally.
\end{abstract}


\maketitle

In a recent paper \cite{Ijjas:2013vea}, we have shown that cosmic microwave background data gathered from the Wilkinson Microwave Anisotropy Probe (WMAP) and the Atacama Cosmology Telescope (ACT) and confirmed by Planck2013 disfavors the simplest inflaton potentials and introduces new difficulties for the paradigm.  
In their response  \cite{Guth:2013sya}, Guth, Kaiser, and Nomura (GKN) countered that cosmic inflation is ``on stronger footing than ever,'' [\textsc{gkn}1]\footnote{Throughout this note, [\textsc{gkn\#}] refers to specific quotes from \cite{Guth:2013sya} that have been reproduced in the Appendix, though we strongly suggest reading \cite{Guth:2013sya} in its entirety.} and Linde \cite{Linde:2014nna} has expressed his support of that view.  What is clear from GKN, though, is that two very different versions of inflation are being discussed.

One is the inflationary paradigm described in textbooks \cite{Mukhanov:2005sc,Dodelson:2003ft}, which we will call {\it classic inflation}. Classic inflation proposes that, beginning from typical initial conditions and assuming a simple inflaton potential with a minimum of fine-tuning, inflation can create exponentially large volumes of space that are generically homogeneous, isotropic and flat, with a nearly scale-invariant spectrum of density and gravitational wave fluctuations that is adiabatic, Gaussian and has generic predictable properties. Implicit in classic inflation is reliance on volume as being the natural measure: {\it e.g.,} even if the probability of obtaining a patch of space with the right initial conditions is small {\it a priori}, the inflated regions occupy an overwhelming volume {\it a posteriori} and so their properties constitute the predictions.

Until now, the problematic issues of classic inflation have been conceptual:  the entropy problem \cite{Penrose:1988mg}, the Liouville problem \cite{Gibbons:2006pa}, the multiverse unpredictability problem \cite{Steinhardt:1982kg,Vilenkin:1983xq,Guth:2000ka}, etc.  Our point in \cite{Ijjas:2013vea} was to show that, even if the conceptual problems are favorably resolved, classic inflation is now disfavored by observations.  It is significant that neither GKN nor Linde dispute these points, as we will detail below [\textsc{gkn2--6}].

Instead, GKN label {\it classic inflation} as "outdated" and, over the course of their paper, they describe an alternative inflationary paradigm that has been developing in recent years and revises the assumptions and goals of inflation, and, as Linde suggests, perhaps of science generally.  This makes clear that a schism has erupted between {\it classic inflation} and what might appropriately be called  {\it postmodern" inflation}.    The two inflationary paradigms are substantially different and should be judged separately.  We will first review the situation for classic inflation, where there is a consensus on its status.  Then, we will describe postmodern inflation and briefly comment on its properties.

\textit{Classic inflation.}   Three independent inputs must be specified to determine predictions of any inflationary scenario, whether classic or postmodern:  the initial conditions, the inflaton potential, and the measure.  The initial conditions refer to the earliest time when classical general relativity begins to be a good approximation for describing cosmic evolution, typically the Planck time. (Here we are assuming for simplicity that inflation is driven by a scalar field slowly rolling down an inflaton potential, but our discussion can be easily generalized to other sources of inflationary energy.) Roughly, the \textit{inflaton potential} determines a family of classical trajectories, some of which do and some of which do not include a long period of inflation; the initial conditions pick out a subset of trajectories; and the measure defines the relative ``weight'' among the subset of trajectories needed to compute the predictions. 
\\ 
%
%
\begin{table*}[t]
\begin{center}
\renewcommand\tabcolsep{0pt}
\renewcommand{\arraystretch}{1}
{\small 
\begin{tabular}{|c|c|c|c|c|}\hline
\rowcolor{PaleSilver}
{\parbox{2.55cm}{\quad}}&
\multicolumn{4}{>{\columncolor{PaleSilver}}l|}{
{\parbox{3.53cm}{\bf{Inflaton Potential}}} + {\parbox{3.3cm}{\bf{Initial Conditions}}} + {\parbox{3.25cm}{\bf{Measure}}} $\Longrightarrow$ {\parbox{3cm}{\bf{Predictions}}}
}\\ \hline
\rowcolor{LightGreen}
&{\parbox{3.75cm}\quad} & {\parbox{3.75cm}\quad} & {\parbox{3.75cm}\quad} & {\parbox{3.75cm}\quad}\\[-.1pt]
\rowcolor{LightGreen}& &&&\\[-.1pt]
\rowcolor{LightGreen}& &&&\\[-.1pt]
\rowcolor{LightGreen}& &&&\\[-.1pt]
\rowcolor{LightGreen}& &&&\\[-.1pt]
\rowcolor{LightGreen}& &&&\\[-.1pt]
\rowcolor{LightGreen}& &&&\\[-.1pt]
\rowcolor{LightGreen}&&&&\\[-.1pt]
\rowcolor{LightGreen} \multirow{-9}{2.1cm}{\bf{Classic inflationary paradigm}}
& \multirow{-9}{3.5cm}{{\bf{Simple --}}\\
Single, continuous stage of inflation governed by potentials with the fewest degrees of freedom, fewest parameters, least tuning.
}  
&\multirow{-9}{3.5cm}{{\bf{Insensitive --}} \\
Inflation transforms typical initial conditions emerging from the big bang into a flat, smooth universe with certain generic properties. \\
}
&\multirow{-10}{3.5cm}{{\bf{Common-sense --}}\\ 
It is more likely to live in an inflated region because inflation exponentially increases volume $\Rightarrow$~measure~=~volume
}
&\multirow{-10}{3.5cm}{{\bf{Generic --}}\\
Based on simplest potentials:\\
- red tilt: n$_S \sim .94-.97$,\\
- large $r \sim .1-.3$*,\\
- negligible f$_{\text{NL}}$,\\
- flatness \& homogeneity}\\[1.5pt]
\hline
\rowcolor{LightCyan} & &&&\\[-.1pt] 
\rowcolor{LightCyan} & &&&\\[-.1pt] 
\rowcolor{LightCyan} & &&&\\[-.1pt]
\rowcolor{LightCyan} & &&&\\[-.1pt] 
\rowcolor{LightCyan} & &&&\\[-.1pt] 
\rowcolor{LightCyan} & &&&\\[-.1pt] 
\rowcolor{LightCyan} & &&&\\[-.1pt] 
\rowcolor{LightCyan} & &&&\\[-.1pt] 
\rowcolor{LightCyan} & &&&\\[-.1pt] 
\rowcolor{LightCyan}& &&&\\[-.1pt] 
\rowcolor{LightCyan}& &&&\\[-.1pt] 
\rowcolor{LightCyan} \multirow{-12}{2.1cm}{\bf{Conceptual problems known prior~to WMAP, ACT \& Planck2013}} & 
\multirow{-17}{3.5cm}{{\bf{Not so simple --}}\\  Even simplest potentials require fine-tuning of parameters to obtain the right amplitude of density fluctuations.}  
&\multirow{-13}{3.5cm}{{\bf{Sensitive --}}\\ 
The initial conditions required to begin inflation are entropically disfavored/exponentially unlikely.  There generically exist more homogeneous and flat solutions without inflation than with.}
& \multirow{-12}{3.5cm}{{\bf{Catastrophic failure~--}} \\
Inflation produces a multiverse in which most of the volume today is inflating and, among  non-inflating volumes (bubbles), Inflation predicts our universe to be exponentially unlikely.
}
&\multirow{-13}{3.5cm}{{\bf{Predictability problem~{--}}}\\
No generic predictions; ``anything can happen and will happen an infinite number of times.'' 
The probability by volume of our observable universe is less than $10^{-10^{55}}$. 
}
\\[1.5pt]
\hline
\rowcolor{LightRed}& &&&\\[-.1pt]
\rowcolor{LightRed}& &&&\\[-.1pt]
\rowcolor{LightRed}& &&&\\[-.1pt]
\rowcolor{LightRed}& &&&\\[-.1pt]
\rowcolor{LightRed}& &&&\\[-.1pt]
\rowcolor{LightRed}& &&&\\[-.1pt]
\rowcolor{LightRed}& &&&\\[-.1pt]
\rowcolor{LightRed}& &&&\\[-.1pt]
\rowcolor{LightRed}& &&&\\[-.1pt]
\rowcolor{LightRed}\multirow{-10}{2.35cm}{\bf{Observational problems after WMAP, ACT \& Planck2013 \cite{Ijjas:2013vea}***}} & 
\multirow{-10}{3.5cm}{{\bf{Unlikeliness problem~--}}\\ 
\textit{Simplest} inflaton potentials  disfavored by data; favored (plateau) potentials require more parameters, more tuning, and produce less inflation.}  
&\multirow{-10}{3.5cm}{{\bf{New initial conditions problem --}}\\ 
Favored plateau potentials require an initially homogeneous patch that is a billion times** larger than required for the simplest inflaton potentials.}
& 
\multirow{-12}{3.5cm}{{\bf{New measure problem --}}\\ 
All favored models predict a multiverse yet data fits predictions assuming no multiverse.
}
&\multirow{-11}{3.5cm}{{\bf{Predictability problem unresolved --}}\\
Potentials favored by data do not avoid the multiverse or the predictability problems above.  Hence, no generic predictions.}
\\
\hline
\end{tabular}}
\end{center}
\label{default}
\caption{Classic Inflation.\newline
\\
*The same arguments used to derive the ``generic'' predictions of tilt, flatness, etc. in \cite{Guth:2013sya}, also predict the tensor-to-scalar ratio to be 10-30\%.
\\  
**A different value is presented in \cite{Guth:2013sya}  because they only consider initial patches that are homogeneous and open, whereas we consider typical patches dominated by various forms energy density such as radiation. 
\\
***Future data can amplify, confirm, or diffuse the three problems introduced in \cite{Ijjas:2013vea}. See Discussion section.
}
\end{table*}

As described in row 1 of Table I, classic inflation is based on assuming simple initial conditions, simple potentials and a simple common-sense measure.   The notion is that, for initial conditions emerging from the big bang, some regions of space have the properties required to undergo a period of accelerated expansion that smoothes and flattens the universe, leaving only tiny perturbations that act as sources of cosmic microwave background fluctuations and seeds for galaxy formation.  Although most regions of space emerging from the big bang may not have the correct conditions to start inflation, this is compensated by the fact that inflation exponentially stretches the volume of the regions that do have the right conditions.   Using volume-weighting as the measure, smooth and flat regions dominate the universe by the end of inflation provided the regions with the correct initial conditions are only modestly rare (though see discussion below). For potentials with a minimum of fields (one) and a minimum of fine-tuning of parameters, there are \textit{generic} inflationary predictions: a spatially flat and homogeneous background universe with a nearly scale-invariant, red-tilted spectrum of primordial density fluctuations ($n_S \sim 0.94-0.97$), significant gravitational-wave signal ($r \sim 0.1-0.3$), and negligible non-Gaussianity ($f_{\textsc{nl}} \sim 0$). 
 
 \textit{Known problems of classic inflation before WMAP, ACT \& Planck2013.}
Conceptual problems with classic inflation have been known for three decades; row 2 of Table I. First, all inflationary \textit{potentials} require orders of magnitude of parameter fine-tuning to yield the observed amplitude of the primordial density fluctuations ($\delta\rho/\rho\sim 10^{-5}$). Second, the probability of a region of space having the right initial conditions to begin inflation is exponentially small \cite{Penrose:1988mg,Gibbons:2006pa}. By standard classical statistical mechanical reasoning, even for simple inflaton potentials, there exist more homogeneous and flat cosmic solutions without a long period of inflation than with inflation \cite{Gibbons:2006pa}. 

The most serious conceptual problem is the \textit{multiverse problem} (sometimes called the \textit{measure problem}) that results from eternal inflation \cite{Steinhardt:1982kg,Vilenkin:1983xq}.  Assuming smooth, classical evolution of the inflaton, inflation comes to an end in a finite time according to when the inflaton reaches the bottom of the inflaton potential.  However, generically, classical evolution is sometimes punctuated by large quantum fluctuations, including ones that kick the inflaton field uphill, far from its expected classical course.  These regions end up undergoing extra inflation that rapidly makes them dominant volumetrically.   In this sense, inflation amplifies rare quantum fluctuations that keep space inflating, leading to eternal inflation.  Continuing along this line of reasoning, there can be multiple quantum jumps of all sorts as the inflaton evolves with time leading to volumes of space (bubbles) with different inflaton trajectories and, consequently, different cosmological properties. For example, some are flat but some not; some have scale-invariant spectrum, some not; etc.  

Ultimately, the result is an eternal multiverse in which ``anything can happen and will happen an infinite number of times.'' [{\sc{gkn}}7] 
What does inflation predict to be the most likely outcome in the multiverse?  In the context of classical inflation, where volume is the natural measure, most volume today is inflating and most non-inflating volume (bubbles) is predicted to be exponentially younger than the observable universe \cite{Linde:1994gy,Guth:2000ka}, [{\sc{gkn}}8].  To be more specific, the volume-weighted prediction is that our observable universe is exponentially unlikely by a factor exceeding $10^{-10^{55}}$ or more! [{\sc{gkn}}9]  Classic inflation is a catastrophic failure by this measure; numerically, it is one of the worst failures in the history of science.  

How has a theory that fails catastrophically continued to survive in scientific discourse?  For the most part, it is because, by ignoring the multiverse and assuming a continuous period of monotonic slow-roll, classic inflation seems to produce predictions that perfectly match observations. The point of \cite{Ijjas:2013vea} was to show that this is no longer the case.

\textit{Problems of classic inflation after WMAP, ACT \& Planck2013.}
WMAP, ACT, and Planck2013 have passed an important milestone.  Like previous experimental groups, they compare their results to an oversimplified version of classic inflation by ignoring the multiverse, as noted above.  For the first time, observational data places pressure on this oversimplified classic inflation.  The new pressure on classic inflation includes the ``unlikeliness problem,'' a new initial conditions problem, and a new measure problem \cite{Ijjas:2013vea}; as summarized in row 3 of Table I. We briefly describe the problems here.

The unlikeliness problem \cite{Ijjas:2013vea} arises because WMAP, ACT \& Planck2013 disfavor the simplest ({\it e.g.}, power-law) inflaton potentials and favors small-field plateau-like potentials. Plateau-like potentials require more tuning, occur for a narrower range of parameters, and produce exponentially less inflation than would be produced by the disfavored power-law potentials\footnote{In counting the maximal number of $e$-folds of inflationary smoothing for a given potential, one should only consider the final inflationary stage during which the density fluctuation $\delta\rho/\rho$ is much less than 1 and exclude inflaton field ranges  where quantum fluctuations dominate classical evolution; see for further discussion Sec.\,III.B of \cite{Ijjas:2013sua}.
}, so it is surprising to find them favored. Furthermore, most energy landscapes with plateau-like inflation paths to the current vacuum also include simple power-law inflation paths to the same vacuum that generate more inflation, so it is exponentially unlikely that the current vacuum resulted from the plateau-like path. Yet this is what WMAP, ACT \& Planck2013 favor. 

The new initial conditions problem arises because the energy density at the beginning of inflation $M_b^4$ is smaller by twelve orders of magnitude in the observationally favored models compared to the simplest inflaton potentials.  In order for inflation to begin, a smooth patch of size $M_b^{-3}$ Hubble volumes (as evaluated at the Planck time in Planck units) is required.  Quantitatively, the observationally favored potentials require an initial smooth patch that is $10^9$ Hubble volumes -- a billion times larger than what is needed to begin inflation for the simplest inflaton potentials.  Since larger smooth patches are exponentially rarer than smaller ones, the favored potentials require comparatively improbable initial conditions.

A third issue that arises due to observations is new challenges for resolving the multiverse measure problem.  For classic inflation, volume-weighting was considered fine for making predictions until the discovery of the multiverse, when it was found that Hubble-sized patches of space like ours are highly improbable.  The challenge for the last three decades has been to find an alternative weighting in the multiverse that will restore the naive volume-weighted predictions.   That program has been unsuccessful to date, so there is no justification for expecting that a small-field plateau potential should produce values of $n_s$, $r$ and $f_{\textsc{nl}}$ that agree precisely with the naive volume-weighted predictions; yet these are the values that Planck2013 has found. 
This imposes a new tight constraint on any solution to the measure problem:  one must seek a clever choice of weighting that can reproduce the naive volume-weighted predictions of classic inflation for plateau-potentials.  However, then there is another twist.  Using the same naive volume-weighting, we have shown in \cite{Ijjas:2013vea} that simple potentials are exponentially favored over the small-field plateau models.  Hence, the solution to the measure problem must mimic naive volume-weighting for some predictions but not for others.  These are new data-imposed restrictions for solving the measure problem. 
%
%
\begin{table*}[htdp]
\begin{center}
\renewcommand\tabcolsep{0pt}
{\small 
\begin{tabular}{|c|c|c|c|c|}\hline
\rowcolor{PaleSilver}
{\parbox{2.25cm}{\quad}}&
\multicolumn{4}{>{\columncolor{PaleSilver}}l|}{
{\parbox{3.53cm}{\bf{Inflaton Potential}}} + {\parbox{3.3cm}{\bf{Initial Conditions}}} + {\parbox{3.25cm}{\bf{Measure}}} $\Longrightarrow$ {\parbox{3cm}{\bf{Predictions}}}
}\\ \hline
\rowcolor{crazy}
&{\parbox{3.75cm}\quad} & {\parbox{3.75cm}\quad} & {\parbox{3.75cm}\quad} & {\parbox{3.75cm}\quad}\\[-.1pt]
\rowcolor{crazy}& &&&\\[-.1pt]
\rowcolor{crazy}& &&&\\[-.1pt]
\rowcolor{crazy}& &&&\\[-.1pt]
\rowcolor{crazy}& &&&\\[-.1pt]
\rowcolor{crazy}& &&&\\[-.1pt]
\rowcolor{crazy}& &&&\\[-.1pt]
\rowcolor{crazy}& &&&\\[-.1pt]
\rowcolor{crazy}& &&&\\[-.1pt]
\rowcolor{crazy}& &&&\\[-.1pt]
\rowcolor{crazy}
\multirow{-12}{2.0cm}{\bf{Postmodern inflationary paradigm}} & 
\multirow{-11}{3.5cm}{{\bf{Complex --}}\\   
with many fields, parameters, dips, minima, and hence many metastable states, leading to multiple phases of inflation [{\sc{gkn}}10-11]
and making eternal inflation unavoidable  [{\sc{gkn}}12]
}  
&\multirow{-15}{3.5cm}{{\bf{Not important --}}\\
in considering validity of inflation; any problems can be compensated by adjusting the measure  {[{\sc{gkn}}19]}}
& 
\multirow{-16}{3.5cm}{{\bf{To be determined --}}\\
from some combination of probability weighting and anthropic selection [\textsc{gkn}13,17,20]}
&\multirow{-12}{3.5cm}{{\bf{Generic --}}\\ 
predictions should generically agree with observations once the right complex potential and combination of measure and anthropic weighting is identified [\textsc{gkn}6,15]}
\\[1.5pt]
\hline
\rowcolor{LightRed}& &&&\\
\rowcolor{LightRed}& &&&\\[-.1pt]
\rowcolor{LightRed}& &&&\\[-.1pt]
\rowcolor{LightRed}& &&&\\[-.1pt]
\rowcolor{LightRed}& &&&\\[-.1pt]
\rowcolor{LightRed}& &&&\\[-.1pt]
\rowcolor{LightRed}& &&&\\[-.1pt]
\rowcolor{LightRed}& &&&\\[-.1pt]
\rowcolor{LightRed}& &&&\\[-.1pt]
\rowcolor{LightRed}\multirow{-10}{2.0cm}{\bf{Problems}} & 
\multirow{-10}{3.5cm}{{\bf{Unpredictability. Part I --}}\\ 
A complex energy landscape allows virtually any outcome and provides no way to determine which inflaton potential form is most likely. [\textsc{gkn}17] 
}  
&\multirow{-12}{3.5cm}{{\bf{Unpredictability. Part II --}}\\
Without knowing initial conditions cannot make predictions even if energy landscape is known. {[{\sc{gkn}}14]}
}& 
\multirow{-10}{3.5cm}{{\bf{Paradigm rests entirely on the measure~--}}\\ 
yet, to date, no successful measure has been proposed and there is no obvious way to solve this problem.  [\textsc{gkn}13]
}
&\multirow{-11}{3.5cm}{{\bf{No predictions -- }}\\
the simplest (volume) measure gives catastrophic results and different landscapes, initial conditions, and measures give different predictions  {[{\sc{gkn}}6]}.  
}\\[1.5pt]
\hline
\end{tabular}}
\end{center}
\label{default}
\caption{Postmodern Inflation.
}
\end{table*}

 \textit{Postmodern inflation.}
From the three new problems we concluded after WMAP, ACT \& Planck2013 that classic inflation is observationally disfavored \cite{Ijjas:2013vea} -- a point which GKN are not disputing [{\sc{gkn}}5]. Instead, they claim that classic inflation must be replaced by a more recent paradigm; that we dub {\it postmodern inflation}. 'Postmodern' is a term used in literature, art, philosophy, architecture, and cultural or literary criticism for approaches that reject the idea of universal truths and, instead, deconstruct traditional viewpoints and focus on relative truths. The term seems to be appropriate to the new inflationary paradigm in which the physical laws and cosmological properties in our observable universe, although apparently uniform, may only be locally valid, with completely different laws and properties in regions outside our horizon and beyond any conceivable causal contact. 

The postmodern approach makes different assumptions about the three inputs used to make inflationary predictions; row 1 of Table II.
\begin{itemize}[label=\tiny$\blacksquare$]  
\item Assuming simple inflaton potentials with a single phase of inflation is ``not at all realistic,'' whereas highly complex potentials with many parameters, tunings, and fields are ``very plausible according to recent ideas in high-energy physics'' [{\sc{gkn}}10--11]. 
The complex potentials inevitably lead to multiple stages of inflation and a multiverse in which anything can happen [{\sc{gkn}}7].  
\item The validity of the postmodern inflationary paradigm cannot be judged on whether it works for typical initial conditions since we do not know what those conditions are [{\sc{gkn}}13].  
Even if the initial conditions are determined some day they will not affect the validity of inflation; rather, the (yet unknown) measure will then be adjusted such that the observed properties of the universe are likely to emerge from those (yet unknown) initial conditions [{\sc{gkn}}14].
\item The volume measure is rejected in favor of complex measures that are to be (re-)adjusted ({\it a posteriori}) to ensure that the predicted outcome agrees with observations. 
\end{itemize}
 
 \textit{Problems of postmodern inflation.} 
Postmodern inflation has its own issues.
One problem arises from allowing highly complex potentials with more parameters than there are observables.  Even if initial conditions were somehow fixed and the multiverse avoided, complex potentials introduce  their own \textit{parameter unpredictability} problem. For example, it has been shown \cite{Ferrara:2013rsa} that a potential with a single field and only three parameters can be designed to fit any cosmological outcome for the standard cosmological observables. If so, then no observation can be said to test the theory. Introducing more degrees of freedom or a complex landscape further exacerbates the situation [{\sc{gkn}}17].
 
A second issue relates to the claim that obtaining inflationary initial conditions following the big bang is unimportant to the validity of the paradigm.  For some cosmologists, this revision will come as somewhat of a shock, since a common justification for introducing inflation is to explain how the current universe can naturally and robustly emerge from a wide range of possible big bang initial conditions.  That is also why several groups have explored the dependence on initial conditions, with some ultimately concluding that the conditions required to have a long period of classic inflation after the universe emerges from the big bang are extremely rare \cite{Penrose:1988mg,Gibbons:2006pa}.  In postmodern inflation, it is conceded that the period of rapid accelerated expansion by itself does not explain how the universe emerged from typical initial conditions.  Ignorance of initial conditions is claimed instead, and the resolution for how the current universe emerged from initial conditions is relegated to the measure, rather than inflation [{\sc{gkn}}14].

Postmodern inflation rests entirely on the measure. It is the measure alone that is supposed to justify the choice of a particular highly complex  potential among exceedingly many. At the same time, the measure is supposed to solve the initial conditions problem, and the very same measure is supposed to regulate infinities in the multiverse and restore predictiveness.  Such a measure does not currently exist 
-- ``a persuasive theory of probabilities in the multiverse has not yet been found'' [\textsc{gkn}6].
Common-sense volume-weighting of classic inflation is declared invalid, but not because there is a fundamental mathematical or logical or intuitive  inconsistency with the volume measure. In fact, the volume measure may work well for some cosmologies \cite{Johnson:2011aa}. Rather, volume-weighting is discarded because it leads to a catastrophic failure  when applied to eternal inflation (see Table I). 

In postmodern inflation, volume-weighting is abandoned in favor of selecting a measure {\it a posteriori} to fit observations. In this approach, the notion of generic predictions is sacrificed. A paradigm that relies on a multiverse in which anything can happen, with initial conditions yet to be determined,  with complex potentials consisting of multiple fields and parameters, and, then, with the freedom to select the measure {\it a posteriori} cannot have generic predictions. 
In fact, observations cannot falsify postmodern inflation -- failure to match observations leads instead to a change of measure [\textsc{gkn}14]. This places postmodern inflationary cosmology squarely outside the domain of normal science.  Linde concurs \cite{Linde:2014nna}, quoting Steven Weinberg \cite{Weinberg:2005fh}, ``Now we may be at a new turning point, a radical change in what we accept as a legitimate foundation for a physical theory."

\textit{Discussion.}
The focus of our original paper \cite{Ijjas:2013vea} was what we call here the  classic inflationary paradigm. We showed that most recent experimental data imposes new challenges by disfavoring the simplest inflaton potentials.  As we emphasized in the conclusion of that paper, the situation is subject to change depending on future data.  For example, suppose that forthcoming analysis of the Planck polarization data will reverse the trend and find $r>0.13$. Suppose further that there remains negligible non-Gaussianity and running of the spectral index and there is no change in the tilt. 
Then, the three observational challenges (row 3 in Table I) posed in \cite{Ijjas:2013vea} disappear (though the conceptual problems in row 2 of Table I would remain).  
On the other hand, finding $r>0.13$ is not sufficient to ease the problems for classic inflation. 
For example, if the fit to the data requires non-negligible non-Gaussianity or a large running of the spectral index, $|\alpha_s | \gg 0.0001$, would be just as bad for classic inflation as an $r$-value below 0.13. Also note that the old problems (row 2 in Table I) remain irrespectively of future experimental data.
 Other scenarios depending on future data are also discussed in \cite{Ijjas:2013vea}. 
 
GKN discount the classic inflationary paradigm as outdated and instead describe an alternative (postmodern) paradigm.  Here, we have made it clear that these are two very different paradigms sharing the same name and being conflated.   Henceforth, it is essential to distinguish the two paradigms; particularly when interpreting experiments. 

Future data has no significance for the postmodern inflationary paradigm because the potential, initial conditions and measure are chosen {\it a posteriori} to match observations, whatever the results. For example, measuring $r>0.13$ or $r<0.13$ or not detecting any gravitational waves at all makes no difference. 

The scientific question we may be facing in the near future is: 
If classic inflation is outdated and a failure, are we willing to accept postmodern inflation, a construct that lies outside of normal science?  Or is it time to seek an alternative cosmological paradigm?

{\it Acknowledgements.} We thank T. Baker, J.-L. Lehners, J. Pollack, D. Spergel, and C. Steinhardt for comments on the manuscript.
This research was partially supported by the U.S. Department of Energy under grant number DE-FG02- 91ER40671 (PJS), by NSF grant AST-0907890 and NASA grants NNX08AL43G and NNA09DBB30A (AL). AI acknowledges the support of the 
European Research Council via the Starting Grant No.~256994. The work of AI is supported in part by a grant from the John Templeton Foundation. The opinions expressed in this publication are those of the authors and do not necessarily reflect the views of the John Templeton Foundation. 
AI thanks the Physics Department of Princeton University for hospitality while this research was completed.
\appendix*
\section{Appendix}
For the reader's convenience we have reproduced specific quotes from Ref.~\cite{Guth:2013sya}, though we urge reading the paper in its entirety.   Citations refer to version \url{arxiv.org/abs/1312.7619v2}.
\begin{itemize}
\item[{[\small\textsc{gkn}\,1]}] Recent experimental evidence, including the impressive measurements with the {\it Planck} satellite of the \textsc{cmb} temperature perturbation spectrum and the strong indication from the \textsc{lhc} that fundamental scalar fields such as the Higgs boson really exist, put inflationary cosmology on a stronger footing than ever. [p.\,8] 
\item[{[\small\textsc{gkn}\,2]}] {\sc isl} further argue that the plateau shape of the low-energy part of the potential is not a consequence of
inflation, but instead is chosen only to fit the {\it Planck} data, a situation which they describe as ``trouble for the [inflationary] paradigm.'' It is of course true that inflation does not determine the shape of the potential, and indeed most inflationary theorists, including us, would consider a $m^2\phi^ 2$ or a $\lambda\phi^4$ potential to be a priori quite plausible for the low-energy part of the potential. [p.\,4]
\item[{[\small\textsc{gkn}\,3]}] We agree that if the observable inflation occurred on a plateau-like potential, eternal inflation seems very likely. It can occur either while the scalar field is at or near the top of the plateau, or in a metastable state that preceded the final stage of inflation. We also agree that this leads to the measure problem: in an infinite multiverse, we do not know how to define probabilities. [p.\,5]
\item[{[\small\textsc{gkn}\,4]}] ... since the measure problem is not fully solved, {\sc isl} are certainly justified in using their intuition to decide that eternal inflation seems unlikely to them. [p.\,5]
\item[{[\small\textsc{gkn}\,5]}] 
If the physical system consisted of a single scalar field $\phi$ which started with random initial conditions at the Planck scale, then {\sc isl}'s argument would be persuasive. [p.\,6]
\item[{[\small\textsc{gkn}\,6]}] 
We agree with Ijjas, Steinhardt, and Loeb that important questions remain. A well-tested theory of physics at the Planck scale [\textit{initial conditions}] remains elusive, as does a full understanding of the primordial singularity and of the conditions that preceded the final phase of inflation within our observable universe [\textit{potential}]. Likewise, although significant progress has been made in recent years, a persuasive theory of probabilities in the multiverse has not yet been found [\textit{measure problem}].[p.\,8] 
\item[{[\small\textsc{gkn}\,7]}]
anything can happen and will happen an infinite number of times [p.\,5]
\item[{[\small\textsc{gkn}\,8]}]
While the proper-time cutoff measure seems intuitive, it has been found to lead to a gross inconsistency with experience, often called the ``youngness problem.'' [p.\,6]
\item[{[\small\textsc{gkn}\,9]}]
Pocket universes as old as $\Delta t = 14$ billion years, for example, are suppressed by a factor such as $e^{-3\Delta t/\tau_{\text{min}}} \sim 10^{-10^{55}}$. [p.\,6]
\item[{[\small\textsc{gkn}\,10]}] Given recent developments in 
high-energy theory (e.g. the revised understanding of the vacuum structure in string theory and the idea that the effective theory below the Planck scale may contain multiple -- often separate -- sectors), we find it very plausible that V ($\phi$) is much more complicated than that, with multiple fields and many local minima. [p.\,3]
\item[{[\small\textsc{gkn}\,11]}] In assessing the criticisms of inflation by ISL, ... most stem from an outdated view
in which a single phase of inflation is assumed 
to persist from the Planck scale to the inflationary scale. None of the quantitative predictions from inflationary cosmology for various observables require such an assumption, nor does such an assumption seem at all realistic in the light of recent developments in high-energy
theory. [p.\,8]
\item[{[\small\textsc{gkn}\,12]}] But once we consider a potential energy function with more than one metastable local minimum -- ... eternal inflation seems unavoidable. [p.\,3]
\item[{[\small\textsc{gkn}\,13]}] ... the measure problem: in an infinite multiverse, we do not know how to define probabilities ... We do not yet know what is the correct method of regularization, or even what physical principles might determine the correct answer. [p.\,5]
\item[{[\small\textsc{gkn}\,14]}] Unlike {\sc isl}, we would view the success or failure of such predictions [of conditions at the Planck scale] not as a test of the inflationary paradigm, but rather as part of our exploration of the measure problem. [p.\,5--6] 
\item[{[\small\textsc{gkn}\,15]}] Anthropic selection effects can then make it plausible that we live in a pocket universe that evolved in this way. [p.\,4] 
\item[{[\small\textsc{gkn}\,16]}] These generic predictions are consequences of simple inflationary models, ... confirmed to good precision, most recently with the {\it Planck} satellite. [p.\,1]
\item[{[\small\textsc{gkn}\,17]}] ... the relative probabilities of the two starting points for the last stage of inflation -- plateau-like or outer wall -- become the issue of complicated dynamics in the multiverse, and we are unable to compute which will dominate with our current knowledge and technology. [p.\,7]
\item[{[\small\textsc{gkn}\,18]}] the possibility that the final stage of inflation was preceded by a bubble nucleation event is at least one way that fine-tuning issues can be avoided. [p.\,3]  
\item[{[\small\textsc{gkn}\,19]}]
We also believe, as a matter of principle, that it is totally inappropriate to judge inflation on how well it fits with anybody's speculative ideas about Planck-scale physics -- physics that is well beyond what is observationally tested. ... and we should similarly not even consider rejecting the inflationary paradigm because it is not yet part of a complete solution to the ultimate mystery of the origin of the universe.  [p.\,2--3] 
\item[{[\small\textsc{gkn}\,20]}] ... important advances have been made in recent years on topics such as eternal inflation, the multiverse and various proposals to define probabilities, and the possible role of anthropic selection effects. [p.\,2]
\end{itemize}

\bibliographystyle{apsrev4-1}

\bibliography{guthrefs}

\end{document}